\documentclass[12pt]{iopart} 

\input{axodraw.sty}

\newcommand{\al}{\alpha}
\newcommand{\be}{\beta}

\newcommand{\VH}{V_{\rm H}}
\newcommand{\VF}{V_{\rm F}}
\newcommand{\VHF}{V_{\rm HF}}
\newcommand{\VB}{V_{\rm BCS}}
\newcommand{\VQH}{V_{\rm QH}}
\newcommand{\Vmx}{V_{\rm mixed}}

\usepackage{theorem,amssymb,amsbsy,latexsym}

\renewcommand{\gg}{g_0}
\newcommand{\vv}{v} 
\newcommand{\nd}{{\vphantom{\dag}}}
\newcommand{\sumdash}{{\mathop{{\sum}'}}}
\newcommand{\cE}{{\cal E}}
\newcommand{\cS}{{\cal S}}
\newcommand{\cZ}{{\cal Z}}
\newcommand{\cY}{{\cal Y}}
\newcommand{\gl}{{\rm gl}}
\newcommand{\SU}{{\rm SU}}
\newcommand{\vsig}{{\mbf \sigma}}
\newcommand{\vS}{{\bf S}}
\newcommand{\vQ}{{\bf Q}}
\newcommand{\vK}{{\bf K}}
\newcommand{\mbf}[1]{\mbox{\boldmath ${#1}$}}
\newcommand{\ee}{{\rm e}}
\newcommand{\ii}{{\rm i}}
\newcommand{\dd}{{\rm d}}
\newcommand{\Om}{\Omega}
\newcommand{\sig}{\sigma}

\newcommand{\vx}{{\bf x}}
\newcommand{\vy}{{\bf y}}
\newcommand{\vk}{{\bf k}}
\newcommand{\vq}{{\bf q}}
\newcommand{\vn}{{\bf n}}
\newcommand{\vnull}{{\bf 0}}

\newcommand{\Ref}[1]{(\ref{#1})}
\newcommand{\eps}{\varepsilon}

\newcommand{\R}{{\mathbb R}}

\newcommand{\eq}{\begin{equation}}
\newcommand{\eqend}{\end{equation}}
\newcommand{\eqa}{\begin{eqnarray}}
\newcommand{\nonueqa}{\begin{eqnarray*}}
\newcommand{\eqaend}{\end{eqnarray}}
\newcommand{\nonueqaend}{\end{eqnarray*}}
\newcommand{\nonu}{\nonumber \\ \nopagebreak}

\begin{document}

\title[2D interacting fermion models]{Exactly solvable models for 2D
interacting fermions\footnote{Contribution to RAQIS'03,
Annecy-le-Vieux, March 2003}}

\author{Edwin Langmann} 

\address{Department of Physics -- Mathematical Physics, Royal Institute of
Technology, AlbaNova -- SCFAB, SE-10691 Stockholm, Sweden}

\ead{langmann@theophys.kth.se} 

\date{\today} 


\begin{abstract} 
I discuss many-body models for interacting fermions in two space
dimensions which can be solved exactly using group theory. The
simplest example is a model of a quantum Hall system: 2D fermions in a
constant magnetic field and a particular non-local 4-point
interaction. It is exactly solvable due to a dynamical symmetry
corresponding to the Lie algebra $\gl_\infty\oplus \gl_\infty$. There
is an algorithm to construct all energy eigenvalues and eigenfunctions
of this model. The latter are, in general, many-body states with
spatial correlations. The model also has a non-trivial zero
temperature phase diagram. I point out that this QH model can be
obtained from a more realistic one using a truncation procedure
generalizing a similar one leading to mean field theory.  Applying
this truncation procedure to other 2D fermion models I obtain various
simplified models of increasing complexity which generalize mean field
theory by taking into account non-trivial correlations but
nevertheless are treatable by exact methods.
\end{abstract}


\newcommand{\Sct}[1]{\section{{#1}}}
\newcommand{\sct}[1]{\subsection{{#1}}}
\newcommand{\ssct}[1]{\subsubsection{{#1}.}}

\newcommand{\figureV}{ \setlength{\unitlength}{1pt}
\mbox{$:c^\dag_{\vk_1,\al} c^\nd_{\vk_2,\al} c^\dag_{\vk_3,\be}
c^\nd_{\vk_4,\be} : \; \, \qquad = $}
\parbox[c]{80pt}{
\begin{picture}(80,130)(-40,-60)
  \ArrowLine(30,0)(0,30)
  \ArrowLine(0,-30)(30,0)
  \ArrowLine(30,0)(60,30)
  \ArrowLine(60,-30)(30,0)
  \put(0,45){\makebox(0,0)[t]{$\vk_1$}}
  \put(0,-35){\makebox(0,0)[t]{$\vk_2$}}
  \put(60,45){\makebox(0,0)[t]{$\vk_3$}}
  \put(60,-35){\makebox(0,0)[t]{$\vk_4$}}
\end{picture}}}

\newcommand{\figureHF}{
\setlength{\unitlength}{1pt}
\parbox[c]{115pt}{
\begin{picture}(80,130)(-40,-60)
  \ArrowLine(30,0)(0,30)
  \ArrowLine(0,-30)(30,0)
  \ArrowLine(30,0)(60,30)
  \ArrowLine(60,-30)(30,0)
  \put(0,45){\makebox(0,0)[t]{$\vk$}}
  \put(0,-35){\makebox(0,0)[t]{$\vk$}}
  \put(60,45){\makebox(0,0)[t]{$\vq$}}
  \put(60,-35){\makebox(0,0)[t]{$\vq$}}
  \put(30,-55){\makebox(0,0)[t]{Hartree terms}}
\end{picture}}
\mbox{$\hspace{8mm}+\hspace{-2mm}$}
\parbox[c]{80pt}{
\begin{picture}(80,130)(-40,-60)
  \ArrowLine(30,0)(0,30)
  \ArrowLine(0,-30)(30,0)
  \ArrowLine(30,0)(60,30)
  \ArrowLine(60,-30)(30,0)
  \put(0,45){\makebox(0,0)[t]{$\vk$}}
  \put(0,-35){\makebox(0,0)[t]{$\vk$}}
  \put(60,45){\makebox(0,0)[t]{$\vq$}}
  \put(60,-35){\makebox(0,0)[t]{$\vq$}}
  \put(30,-55){\makebox(0,0)[t]{Fock terms}}
\end{picture}}}

\newcommand{\figureBCS}{
\setlength{\unitlength}{1pt}
\parbox[c]{115pt}{
\begin{picture}(80,130)(-40,-60)
  \ArrowLine(30,0)(0,30)
  \ArrowLine(0,-30)(30,0)
  \ArrowLine(30,0)(60,30)
  \ArrowLine(60,-30)(30,0)
  \put(0,45){\makebox(0,0)[t]{$(\vq,\uparrow)$}}
  \put(0,-35){\makebox(0,0)[t]{$(\vk,\uparrow)$}}
  \put(60,45){\makebox(0,0)[t]{$(-\vq,\downarrow)$}}
  \put(60,-35){\makebox(0,0)[t]{$(-\vk,\downarrow)$}}
  \put(30,-55){\makebox(0,0)[t]{BCS terms}}
\end{picture}}}

\newcommand{\figuremx}{
\setlength{\unitlength}{1pt}
\parbox[c]{115pt}{
\begin{picture}(80,130)(-40,-60)
  \ArrowLine(30,0)(0,30)
  \ArrowLine(0,-30)(30,0)
  \ArrowLine(30,0)(60,30)
  \ArrowLine(60,-30)(30,0)
  \put(0,45){\makebox(0,0)[t]{$(k_x,q_y)$}}
  \put(0,-35){\makebox(0,0)[t]{$(k_x,k_y)$}}
  \put(60,45){\makebox(0,0)[t]{$(q_x,k_y)$}}
  \put(60,-35){\makebox(0,0)[t]{$(q_x,q_y)$}}
  \put(30,-55){\makebox(0,0)[t]{mixed terms I}}
\end{picture}}
\mbox{$\hspace{8mm}+\hspace{-2mm}$}
\parbox[c]{80pt}{
\begin{picture}(80,130)(-40,-60)
  \ArrowLine(30,0)(0,30)
  \ArrowLine(0,-30)(30,0)
  \ArrowLine(30,0)(60,30)
  \ArrowLine(60,-30)(30,0)
  \put(0,45){\makebox(0,0)[t]{$(q_x,k_y)$}}
  \put(0,-35){\makebox(0,0)[t]{$(k_x,k_y)$}}
  \put(60,45){\makebox(0,0)[t]{$(k_x,q_y)$}}
  \put(60,-35){\makebox(0,0)[t]{$(q_x,q_y)$}}
  \put(30,-55){\makebox(0,0)[t]{mixed terms II}}
\end{picture}}}

\Sct{Introduction} In this paper I present a class of exactly solvable
many-body models which describe two dimensional (2D) interacting
fermion systems with spatial correlations.  By `exactly solvable
model' I mean that there is an algorithm to construct all eigenvalues
and eigenstates of the many-body Hamiltonian explicitly. By
`correlated' I mean that the eigenstates, and in particular the
groundstate, are not Slater states in general. I introduce these
models by discussing a simple example which I call {\em QH
model}. This is a model of 2D electrons in a magnetic field and
interacting with a particular 4-point interaction \cite{EL}. I then
propose a novel interpretation of this model allowing for natural
generalizations and leading us to a large class of exactly solvable
models of interacting fermions. As I will explain, all these models
can be obtained from more realistic ones by a truncation procedure
which generalizes a similar one leading to mean-field theory.  As a
motivation, I will discuss two such models which played an important
role in condensed matter physics: the BCS model which solved the
problem of superconductivity, and the Hartree-Fock model providing a
basis for understanding magnetism in metals. These models can be
obtained from a more general one by discarding all but particular
kinds of terms in the interactions. The Hartree-Fock model is obtained
by keeping only the so-called Hartree- and Fock terms, i.e.\ the terms
describing two-body scattering processes where the fermion momenta are
left unchanged and are exchanged, respectively; see the Fig.\ in
\Ref{FigHF} below.  In 2D there are also mixed scattering terms which
are Hartree-like in the $x$- and Fock-like in the $y$-component of the
momenta and {\em vice versa}; see Fig.\ \Ref{Figmx}. I observe that
the interaction in the QH model is a sum of such mixed terms, and I
propose generalized models where, in addition to Hartree- and Fock
terms, also mixed terms are included. These models are interesting
since they include non-trivial spatial correlations but still can be
treated by exact methods.


I only mention in passing some physics motivation (and refer readers
interested in more details to a review \cite{review}). There are
several fascinating phenomena which have been discovered in complex
transition metal oxides in recent years. A prominent example is high
temperature superconductivity. 2D models of Hubbard-type are generally
believed to account for these phenomena, but these models have been
found to be very difficult: despite of much work they are still poorly
understood. I therefore believe that the simplified models proposed
here will be useful in this context.

It is important to note that the exact solution of Hartree-Fock type
models is equivalent to a mean field approximation of a more
complicated model (I will discuss this in more detail in Section
\ref{mft}), and the same is true for the BCS model.\footnote{I believe
this is the reason why these models are no longer widely known: there
are other, more efficient methods to derive mean field theory.}  We
thus propose that our 2D models provide a generalized mean field
theory taking into account non-trivial spatial correlations. The
latter is somewhat more complicated but still can be computed by exact
methods. To be more specific: Standard mean field theory corresponds
to minimizing a functional depending on a small number of variation
parameters, whereas the generalized mean field theory is given by a
matrix model. E.g.\ the grand canonical partition function of the QH
model can be written as a hermitean matrix integral
\eq \cZ = \int\dd^{L^2} X\, \ee^{-L F(X) } \eqend
with a function $F$ only depending on the eigenvalues of the $L\times
L$ hermitean matrix $X$, and the thermodynamic limit corresponds to
$L\to\infty$. Much is known how to treat such matrix models exactly
(see e.g.\ \cite{Mehta} for an introduction to this field). I believe
that it is possible to find exact expressions for these partition
functions.

The simplified models discussed in this paper are such that
eigenstates and eigenvalues can be found exactly using group theory.
Models of this kind are standard in nuclear physics \cite{NuclPhys}
but have been used surprisingly little in condensed matter physics.
An exact solution of a BCS model \cite{BCSorig} found by Richardson a
long time ago \cite{BCS} went unnoticed in condensed matter physics
until recently \cite{BCSnew}. This model and some recent
generalizations \cite{BCSnew1} are similar to the models proposed here
in that all of them can be solved using group theory. In fact, the
model solved in \cite{BCS} is similar to what we call Hartree-models
in that the corresponding Lie groups are finite dimensional (SU(2),
e.g.), whereas our `mixed' models are related to infinite dimensional
Lie groups.

The plan of the rest of this paper is as follows. In the next section
we discuss the QH model. I point out various remarkable properties of
this model and present its solution obtained in \cite{EL}.  Section 3
contains a discussion of general interacting fermion systems and the
truncation procedure leading to simplified models which can be solved
exactly. The 2D Hubbard model is a special case where our truncation
procedure leads to particular simple models.  We end with a short
summary in Section 4.

An outline of this article appeared in \cite{EL0}.

\Sct{The Quantum Hall model} In this Section I discuss a model of
interacting fermions proposed and solved in \cite{EL}. 

\sct{Definition and physical interpretation} \label{s11} We start with
the Landau Hamiltonian
\eq
\label{hB}
h_B = (-\ii \partial_x + By)^2 + (-\ii \partial_y - Bx)^2 
\eqend
describing the motion of a charged particle in the $(x,y)$-plane with
a constant magnetic field of magnitude $B>0$ perpendicular to this
plane. We choose units such that the mass and charge of the electron
are $1/2$ and $2$, respectively.  A microscopic model for a quantum
Hall system\footnote{I do {\em not} write `quantum Hall {\em effect}
model' since, firstly, such a model should also include impurities
and, secondly, I do not know if the interactions I propose below can
account for the QHE. Note also that I ignore spin.} is given by the
many-body Hamiltonian $H=H_0 + V$ where 
\eq
\label{HB} 
H_0 = \int_{\R^2} \dd^2\vx \, \psi^\dag(\vx) \bigl( h_B - \mu \bigr) \psi(\vx), 
\eqend
with fermion creation- and annihilation operators $\psi$ and
$\psi^\dag$ obeying the usual canonical anticommutator relations, $\{
\psi(\vx),\psi^\dag(\vx')\} = \delta^2(\vx-\vx')$ etc.; $\vx \equiv
(x,y)$, and the real parameter $\mu$ is a chemical potential.  The
interactions $V$ is of the following two-body type,
\eqa \label{Vx} V = \int 
\dd^2\vx_1 \ldots \dd^2\vx_4 \,
v(\vx_1,\ldots,\vx_4) : \psi^\dag(\vx_1) \psi(\vx_2) \psi^\dag(\vx_3)
\psi(\vx_4) : \eqaend
with the dots indicating normal ordering as usual.\footnote{
$:\psi^\dag(1)\psi(2) \psi^\dag(3) \psi(1): \; =
\psi^\dag(3)\psi^\dag(1)\psi(2)\psi(4)$} The interaction vertices $v$
of main physical interest for quantum Hall physics are as follows, 
\eq
v(\vx_1,\ldots,\vx_4) = \delta^2(\vx_1-\vx_2)
\delta^2(\vx_3-\vx_4) W(\vx_1-\vx_3) 
\eqend
with $W$ the two-body potential. For example, $W(\vx) = const./|\vx|$
corresponds to a 3D Coulomb repulsion, but this choice leads to a
model which cannot be analyzed without approximations. One might hope
to obtain a simpler model by taking a fully local interaction $W(\vx)
= g\delta^2(\vx-\vy)$, but this model is trivial since the fully local
interaction vanishes due to the Pauli principle, $V=0$.

The exactly solvable QH model can be obtained as a deformation of a
fully local interaction which is such that it becomes non-trivial.  In
Fourier space this deformed interaction is given by
\eqa
\label{Vstar} 
\VQH = \frac{\gg}{(2\pi)^2} \int 
\dd^2\vk_1 \ldots \dd^2\vk_4 \,
\delta^2(\vk_1-\vk_2+\vk_3-\vk_4) \times \nonu \qquad \qquad \qquad \times  
\ee^{-\ii \theta ( \vk_1\wedge \vk_2
+ \vk_3 \wedge \vk_4) } \, : \hat \psi^\dag(\vk_1) \hat \psi(\vk_2)
\hat \psi^\dag(\vk_3) \hat \psi(\vk_4) : \eqaend
with the hat indicating Fourier transform, $\vk\wedge \vq = k_x q_y
-k_y q_x$, and $\theta$ the deformation parameters (i.e.\ $\theta=0$
corresponds to the fully local interaction). At first sight this
interaction might appear somewhat unusual, but it has several
remarkable mathematical properties which I will discuss below.  These
mathematical properties were my original motivation to consider this
model \cite{EL}. In the next Section I will discuss other interacting
fermion models which suggest an alternative physical
interpretation. This discussion will lead us to various interesting
generalizations of this QH model which, as I believe, are promising
candidates for understanding 2D correlated fermion in a precise
mathematical framework.

\sct{Mathematical properties} The QH model defined above has various
interesting mathematical properties.  Firstly, the interaction in
\Ref{Vstar} can be obtained from the fully local interaction by
replacing the pointwise product of fields by the so-called Groenewold-Moyal star
product,
\eq
\label{Vstar1}
\VQH = \gg\int_{\R^2} \dd^2\vx\,  
: \psi^\dag(\vx) \star \psi(\vx) \star \psi^\dag(\vx)\star \psi(\vx) : 
\eqend
with $\star$ the associative product defined such that $x\star y = xy
- \ii\theta$, $y\star x = xy + \ii\theta$, $x\star x=x^2$ and $y\star
y=y^2$. Indeed, from these rules one can deduce that the
$\star$-product of two plane waves is $\ee^{\ii\vk\cdot\vx}\star
\ee^{\ii\vq\cdot\vx} =
\ee^{\ii(\vk+\vq)\vx}\ee^{-\ii\theta\vk\wedge\vq}$, and with that one
can compute Eq.\ \Ref{Vstar1} by inserting the Fourier transforms of
the fields and obtain \Ref{Vstar}. As can be seen by searching the
{\em arXive}, field theories with such $\star$-interactions received
much attention in the particle physics literature theory recently
(they are usually referred to as `noncommutative field
theories'). Secondly, the interaction $\VQH$ has the remarkable
property that it looks the same in Fourier- and position space:
Computing the inverse Fourier transformation one finds that $\VQH$ can
be written as in Eq.\ \Ref{Vx} with the interaction vertex \cite{LS}
\eq
v(\vx_1,\ldots,\vx_4) = \frac{\gg}{(2\pi \theta)^2} 
\delta^2(\vx_1-\vx_2+\vx_3-\vx_4) \ee^{-\ii ( \vx_1\wedge \vx_2
+ \vx_3 \wedge \vx_4)/\theta} .  
\eqend
Since the Landau Hamiltonian obviously looks the same in position- and
Fourier space as well, the Hamiltonian $H=H_0+\VQH$ is invariant under
the following duality transformation \cite{LS},
\eqa
\psi(\vx) &\to& \tilde \psi(\vx) = B \hat \psi(B\vx) \nonu
B&\to& \tilde B = B\nonu
\theta&\to& \tilde \theta = - 1/(\theta B^2)\nonu
\gg&\to& \tilde \gg = \gg/|B\theta| .  
\eqaend
Note that this duality transformation involves Fourier transformation,
$\psi(\vx)\to \hat \psi(\vk)$, followed by a re-scaling of the Fourier
variable to correct for the length dimensions, $\vk\to \vk/B \equiv
\vx$ (since $B$ has the dimension $length^{-2}$). Thirdly, as I will
discuss in more detail below, this model is exactly solvable for
$\theta=\pm B^{-1}$ \cite{EL}. It is interesting to note that these
exactly solvable cases are mapped onto each other under the duality
transformation.

We can actually give the solution of a more general model where $h_B$
in Eq.\ \Ref{HB} is replaced by $(1-a) h_B + a h_{-B}$ with $0\leq
a\leq 1$. This generalization means that we allow for a confining
electric potential $\propto (x^2+y^2)$ in addition to a magnetic
field.

\sct{Integrability} \label{91} In the following we assume $\theta=1/B$.

As discussed in most quantum mechanics textbooks, the eigenfunctions
$\phi_{\ell m}(\vx)$ of the Landau Hamiltonian in Eq.\ \Ref{hB} are
labeled by two positive integers $\ell,m$ with eigenvalues depending
only on one of these quantum numbers, $h_B\phi_{\ell m} = e_\ell
\phi_{\ell m}$ where $e_\ell= 4B(\ell-\frac12)$ (this is the
well-known degeneracy of the Landau problem). The quantum numbers
$\ell$ and $m$ can be interchanged either by complex conjugation,
$\phi_{\ell m}(\vx)^* = \phi_{m \ell}(\vx)$, or by changing the
direction of the magnetic field, $h_{-B}\phi_{\ell m} = e_m
\phi_{\ell m}$. Note that changing the sign of the magnetic field is
equivalent to a parity transformation, $\vx\to -\vx$.

These eigenfunctions provide a complete orthonormal basis in the
1-particle Hilbert space $L^2(\R^2)$, and thus the many-body
Hamiltonian $H_0$ in Eq.\ \Ref{HB} can be diagonalized by expanding the
fermion field operators in this Landau basis,
$\psi(\vx) = \sum_{\ell,m} c_{\ell m} \phi_{\ell m}(\vx) $
and similarly for $\psi^\dag(\vx)$. This yields $H_0 = \sum_{\ell,m}
(e^\nd_\ell -\mu) c^\dag_{\ell m} c^\nd_{\ell m}$.
Since the $\phi_{\ell m}$ are
common eigenfunctions of $h_B$ and $h_{-B}$, we can actually
generalize the model by replacing $h_B\to (1-a) h_B + a h_{-B}$ and
still obtain a diagonal many-body Hamiltonian, 
\eq \label{H011} 
H_0 = \sum_{\ell,m}(E^\nd_m + \tilde E^\nd_\ell) c^\dag_{\ell m}
c^\nd_{\ell m} \eqend
with $\tilde E_\ell\equiv (1-a)e_\ell$ and $E_m \equiv a e_m - \mu$.
The fermion field operators $c^\dag_{\ell m}$ and $c^\nd_{\ell m}$
obey the canonical commutation relations $\{c^\nd_{\ell
m},c^\dag_{\ell'm'} \}=\delta_{\ell,\ell'}\delta_{m,m'}$ etc. The
model is defined on a Fermion Fock space with a vacuum $\Omega$
such that $c_{\ell m}\Omega =0$ for all $\ell,m$.

The crucial fact making the QH model `special' are the following
remarkable $\star$-product relations of the Landau eigenfunctions
\cite{EL},
\eq
\phi_{\ell m}\star \phi_{\ell' m'} = \sqrt{\frac{B}{4\pi}} \, \delta_{m,\ell'} \phi_{\ell m'} 
\quad \mbox{ for } \theta=\frac1B. 
\eqend
Using that and other properties of the Landau eigenfunctions, a
straightforward computation brings the interaction in \Ref{Vstar1} to
the following simple form,
\eq
\label{VQH1} 
\VQH = g  \sum : c^\dag_{m\ell}c^\nd_{m\ell'} c^\dag_{m' \ell'}c^\nd_{m' \ell} : 
\eqend
where $g = \gg B/4\pi$. 

It is interesting to note that the QH Hamiltonian can be written in
terms of the following fermion bilinears,
\eq
\label{rho}
\rho^\nd_{m m'} = \sum_\ell c^\dag_{\ell m}c^\nd_{\ell m'} , \quad 
\tilde \rho^\nd_{\ell \ell'} = \sum_m c^\dag_{\ell m}c^\nd_{\ell' m} .
\eqend
This allows us to write 
\eq 
\label{corr}
H_0 = \sum_{m} E_m \rho_{m m} +\tilde E_m \tilde \rho_{m m}
\eqend
and
\eq
\label{123}
\VQH =  g \sum_{\ell,m} : \rho_{m \ell}\rho_{\ell m}: = 
-g\sum_{m, \ell} : \tilde \rho_{\ell m}\tilde \rho_{m \ell}: . 
\eqend
If we regard the operators $\rho_{\ell m}$ as elements of an infinite
matrix $\rho$ and similarly for $\tilde \rho$ we can write this in the
following suggestive matrix notation,
\eq 
H_0 = \Tr(E\rho + \tilde E\tilde\rho),\quad \VQH = g:\Tr(\rho^2): = -g:\Tr(\tilde \rho^2): 
\eqend
with $E={\rm diag}(E_1,E_2\ldots)$ and similarly for $\tilde E$. 

Using the fermion anticommutator relations we derive the following
commutator relations
\eq [\rho_{\ell m}, \rho_{\ell' m'}] = \delta_{m,\ell'}\rho_{\ell m'}
- \delta_{m',\ell}\rho_{\ell' m} \: \eqend 
and similarly for the $\tilde\rho$'s, and $[\rho_{\ell m}, \tilde
\rho_{\ell' m'}]=0$. Moreover,
\eq
\rho_{\ell m}^\dag = \rho_{m\ell}^\nd 
\eqend
and similarly for the $\tilde\rho$'s. We thus see that {\em the
operators $\rho$ and $\tilde\rho$ represent the Lie algebra
$\gl_\infty\oplus \gl_\infty$}.  It is not difficult to check that the
interaction $\VQH$ represents a Casimir element in this Lie algebra
(i.e.\ it commutes with `everything') and the free Hamiltonian $H_0$
represents an element in the Cartan subalgebra (i.e.\ is a sum of
mutually commuting terms). However, the Lie algebra
$\gl_\infty\oplus\gl_\infty$ does {\em not} commute with $H$: one can
check the following commutator relations, ${[}H,\rho_{\ell \ell'}{]}=
(E_\ell-E_{\ell'})\rho_{\ell \ell'}$, and similarly for $\tilde\rho$
and $\tilde E$. Thus this Lie algebra defines a dynamical symmetry of
the QH model. This suggests that the model is integrable. We will show
this by presenting an algorithm to construct all eigenstates and
eigenvalues of $H$ below.

\sct{Solution} \label{92} A complete orthonormal basis in the Fock
space of the model is given by the states
\eq
\label{vN} 
|N \rangle = c^\dag_{\ell_1 m_1}c^\dag_{\ell_2 m_2}\cdots
c^\dag_{\ell_N m_N}\Omega, \eqend
with $N=0,1,2,\ldots$ the fermion number.  We find it convenient to
introduce a regularization by restricting the quantum numbers to
$\ell_j,m_j=1,2\ldots,L$ with a finite cut-off $L$. This
has a natural physical interpretation: we add a box-like potential
confining the QH system to the disc $x^2+y^2\leq
const. \, L^2/B$. This reduces the dynamical symmetry to
$\gl_L\oplus \gl_L$, and the fermion Fock space becomes
finite dimensional, dim$=2^{L^2}$.

Straightforward computations using the canonical anticommutator
relations show that the interaction acts on these states as follows,
\eq
:\Tr(\rho^2): |N \rangle = 
2 \sum_{1\leq i<j\leq N} T_{(ij)} |N \rangle \equiv C_N |N \rangle 
\eqend
with $T_{(ij)}$ transpositions interchanging $m_i$ and $m_j$, 
\eqa
\label{Tt}
T_{(ij)}  |N \rangle = 
c^\dag_{\ell_1 m_1} 
\ldots c^\dag_{\ell_i m_j}\ldots
c^\dag_{\ell_j m_i}\cdots c^\dag_{\ell_N m_N}\Om ,\quad i<j. 
\eqaend 
Note that $T$ defines a representation of the permutation group $S_N$,
and $C_N = \sum_{i<j} T_{(ij)}$ commutes with all permutations
$T_P$. All $|N\rangle$ are, of course, eigenstates of the free part of
the QH Hamiltonian, $H_0 |N \rangle = \cE_0 |N \rangle$ with $\cE_0 =
\sum_{j=1}^N (\tilde E_{\ell_j} + E_{m_j})$, and due to our special
form of $\cE_0$ this free part commutes with all permutations
$T_P$. Thus the eigenstates of $H$ can be constructed by solving the
following eigenvalue equation,
\eq
\label{group} 
C_N \sum_{P\in S_N} a_P T_P = \vv \sum_{P\in S_N} a_P T_P .  
\eqend
{\it Each solution of \Ref{group} provides and eigenstate $\psi$ and
corresponding eigenvalue $\cE$ of the QH Hamiltonian $H$ as follows,
\eq
\label{psi}
\psi=\sum_P a_P T_P|N\rangle \; \; \mbox{ and } \; \; \cE=\sum_{j=1}^N
(E_{\ell_j} + \tilde E_{m_j} ) + 2 g \vv . \eqend
}Eq.\ \Ref{group} can be solved by constructing and diagonalizing
$N!\times N!$ matrices representing $C_N$. To appreciate the
difficulty of this problem it is instructive to do this for
$N=2,3$. Obviously such brute-force approach is restricted to rather
small $N$. Fortunately for us, Eq.\ \Ref{group} corresponds to a
classical problem in group theory which was solved a long time ago
(the group theory results used in the following are discussed in most
advanced group theory books; for more details and specific references
see \cite{EL}): The representation $T$ is (essentially) the regular
representation, and solving Eq.\ \Ref{group} amounts to decomposing
$T$ into irreps. The irreps of $S_N$ are well-known. They are
one-to-one correspondence with partitions
$[\lambda]=[\lambda_1,\lambda_2,\ldots,\lambda_L]$ of $N$ where the
$\lambda_j$'s are integers summing up to $N$ and obeying
$\lambda_1\geq \lambda_2\geq \cdots\geq \lambda_L\geq 1$. The possible
eigenvalues $\vv$ in Eq.\ \Ref{group} are given by the value of the
class operator $C_N$ in the irreps $[\lambda]$,
\eq \vv_{[\lambda]} = \sum_{i=1}^L \frac12\lambda_i(\lambda_i+1) -
\sum_{i=1}^L i \lambda_i , \eqend
and its multiplicity is the square of the dimension the irreps
$[\lambda]$. E.g. for $N=4$ the eigenvalues $c$ of $C_4$ are
$6(1),2(9),0(4),-2(9),-6(1)$ with the numbers in the parenthesis
giving the multiplicities. The corresponding $\sum_P a_P T_P$ is equal
to the Young operator $\cY_{[\lambda]}$ associated with
$[\lambda]$. This gives the following explicit recipe for
constructing the eigenstates associated with $[\lambda]$: {\it For
sets of $K\leq N$ distinct integers $(i_1,i_2,\ldots,i_K)$, $1\leq
i_j\leq N$ , define (anti-)symmetrizers $S^\pm_{i_1,i_2,\ldots,i_K}=
\sum_{P\in S_K}(\pm 1)^{|P|} T_P^{i_1,i_2,\ldots,i_K}$ acting on
states $|N\rangle$ as described above, with the integers $i_j$
indicating the positions of the $m_{i_j}$ on which the permutations
act. Write the integers $1,2,\ldots,N$ into the boxes of the Young
tableaux associated with $[\lambda]$, take for each row one
symmetrizer $S^+$ with indices given by the numbers in the boxes of
the row, and similarly for each column take an antisymmetrizer. Then
\eq
\psi = \cY_{[\lambda]}|N\rangle \equiv \prod \cS^+_{\rm rows}
\prod \cS^-_{\rm columns} |N\rangle .
\eqend
}
For example, the 10-particle eigenstates associated with
$[\lambda]=[5,3,2]$ are
\eq
\label{psi1}
\psi =\cS^+_{1,2,3,4,5} \cS^+_{6,7,8} \cS^+_{9,10} \cS^-_{1,6,9}
\cS^-_{2,7,10} \cS^-_{3,8}|N=10\rangle, \eqend
and $\cE=\sum_{j=1}^{10} (E_{\ell_j} + \tilde E_{k_j}) + 14 g$ are the
corresponding eigenvalues. This gives all possible eigenvalues and
corresponding eigenstates of the QH model. It is worth mentioning that
there is another representation $\tilde T$ of the permutation group
$S_N$ acting on the indices $\ell_j$ of the states $|N\rangle$ in
\Ref{vN}. The two representations are, however, related: interchanging
$m_i\leftrightarrow m_j$ and $\ell_i\leftrightarrow \ell_j$ together is the
same as interchanging two fermions and only gives a minus sign,
and thus $T_P \tilde T_P=(-1)^{|P|}$ for all $P\in S_N$.

It is interesting to note that the model remains solvable if we add to
$H$ the following interaction terms,
\eq 
\label{VQH2}
\VH= \sum v_{\ell m} \rho_{\ell \ell} \rho_{m m} + 
\tilde v_{\ell m} \tilde \rho_{\ell \ell}\tilde \rho_{m m} + w_{\ell m} 
\rho_{\ell \ell} \tilde \rho_{m m} \eqend
with $\tilde v, v$ and $w$ arbitrary parameters: the eigenstates
$\psi$ in Eq.\ \Ref{psi} remain the same, and the eigenvalues $\cE$
are changed by adding $\sum_{i,j}( v_{m_i m_j} + \tilde v_{\ell_i
\ell_j} + w_{m_i \ell_j})$. We will come back to this in Section
\ref{s22}.

\sct{Physical properties} We found all energy eigenstates and
eigenstates of the QH model.  However, this is only a first step to
understand the physics described by this model: much work remains to
be done. In this Section I describe a few results and mention
interesting open problems.

\ssct{Correlations} The $|N\rangle$ in Eq.\ \Ref{vN} are Slater
states: they are eigenstates of the free Hamiltonian in Eq.\
\Ref{H011} and describe fermions without any correlations. In the next
Section we will discuss the so-called Hartree-like models. These are
exactly solvable models with interactions but such that the
eigenstates are still Slater states: the interaction only modifies the
energy eigenvalues but the eigenstates remain uncorrelated. The QH
model solved above is remarkable in that its eigenstates are
particular linear combinations of Slater states. These eigenstates
$\psi$ still are characterized by $N$ pairs of quantum numbers
$(\ell_j,m_j)$, but they are complicated linear combinations of Slater
states $|N\rangle$ where the $\ell_j$'s and $m_j$'s are distributed
over the fermions in many different ways.  The particular
distributions which yield eigenstates are labeled by partitions
$[\lambda]$, and the energy of the state depends not only on the
quantum numbers $(\ell_j,m_j)$ but also on $[\lambda]$. It is
impossible, in general, to write these eigenstates $\psi$ as Slater
states. It would be very interesting to compute 4-points Greens
functions and thus understand the correlations quantitatively.

\ssct{Ground state and partition function} A first step in
understanding the physical properties of the QH model is to determine
the ground state at fixed fermions density $\nu = N/L$ and in the
thermodynamic limit $L\to \infty$. For that one needs to find the
eigenstate $\psi$ at fixed, large particle numbers $N$ such that the
corresponding eigenvalue $\cE$ is minimal. This is a non-trivial
problem whose solution I only know in one simple, special case.

The main difficulty in determining the groundstate of the QH model
from the solution given above is that we actually found too many
eigenstates: For each set of $N$ quantum numbers $(\ell_j,m_j)$ we
constructed $N!$ eigenstates, but these can be linearly independent
only if all $\ell_j$'s and all $m_j$'s are distinct. For example, if
all $m_j$'s in $|N\rangle$ Eq. \Ref{vN} are the same, then $|N\rangle$
is only non-zero if all $\ell_j$'s are distinct (Pauli
principle). Moreover, in this case all permutations act trivially,
$T_P |N\rangle = |N\rangle$, and thus there is only one non-zero
eigenstate, namely $\psi=|N\rangle$, corresponding to $[\lambda]=[N]$
(the Young tableaux with only one row). The opposite extreme is when
all $\ell_j$'s are the same. In this case all $m_j$'s need to be
distinct, and the only non-zero eigenstate corresponds to $[\lambda] =
[1^N]$ (the Young tableaux with only one column).  In general,
degeneracies of the $\ell_j$'s and $m_j$'s will reduce the
multiplicities of the eigenstates given above. Many of these
multiplicities are zero due to the Pauli principle, and to find the
minimum energy eigenvalue with non-zero multiplicity can be difficult.
This seemingly technical point has important physical implications: it
can lead to a frustration in the system, and thus the groundstate can
change drastically as the model parameters are varied. Below I will
illustrate this by discussing the ground state of the QH model as a
function of $g/B$.

More generally one would like to compute the grand canonical partition
function $\cZ=\Tr\exp(-\beta H)$. 
The computation of $\cZ$ seems like a highly non-trivial problem, but
I believe it has a beautiful solution, at least for the case where
$\tilde E=0$: as already mentioned, one can represent $\cZ$ as a
matrix model with an external field $E$, and even though the resulting
matrix model does not seem to be a `standard' one it seems computeable
(it generalizes a matrix model recently solved in \cite{LSZ}). I now
sketch how to represent $\cZ$ as a matrix model.  We start with a
standard representation of $\cZ$ as a fermion functional integral,
i.e.\ an integral over $\tau$-dependent Grassmann numbers where $0\leq
\tau\leq \beta$ is the Matsubara time; see e.g.\ \cite{NO}. One then
introduces a Hubbard-Stratonivich field $Y(\tau)$,
$0\leq\tau\leq\beta$, which allows to integrate out the fermions. The
fermion integral yields a functional determinant $\det(E\otimes I +
I\otimes[\partial_\tau - \ii Y(\tau)])$.  The field $Y$ are hermitean
$L\times L$ matrices and periodic, $Y(0)=Y(\beta)$. Thus $\cZ$ becomes
a hermitean matrix path integral.  One then can change variables to $X
= U(\tau)^{-1} [\ii\partial_\tau + Y(\tau)] U(\tau)$ with $U(\tau)$
unitary matrices and periodic and $X$ a hermitean constant matrix
(independent of $\tau$). The integral over $U(\tau)$ can be done, and
thus one can represent $\cZ$ as an integral over the hermitean
matrices $X$ in an external field $E$. I believe it is possible to
evaluate this matrix integral in the large-$L$ limit by generalizing
the computation of a similar integral in \cite{LSZ}.

\ssct{Zero temperature phase diagram of the QH model} \label{LR} I now
discuss the ground state of the QH model
\eq H = \sum_{m\geq 1} 4 B (m-1) + g \sum_{\ell,m} : \rho_{\ell m}
\rho_{m \ell} : \eqend
as a function of the coupling parameter $g/B$ ($B>0$, $g$ arbitrary
real, and we set $\mu=1/2$ for convenience).  As discussed, the energy
eigenvalues are sums of two terms: the kinetic energy $\cE_0 =
\sum_{j=1}^N 4 B (m_j-1) $, and the energy $\cE_{\rm
corr}=2g\vv_{[\lambda]}$ due to the interaction. We consider the case
$N<L$, i.e.\ the fermion filling factor $\nu=N/L$ is less than 1. Then
the kinetic energy is minimal if all fermions have the same quantum
number $m_j=1$. In this case $|N\rangle$ in Eq.\ \Ref{vN} is also an
eigenstate of the interaction $\VQH$, and the corresponding eigenvalue
is $\cE_{\rm corr} = gN(N-1)$. This obviously leads to a minimal total
energy $\cE_{[N]}= gN(N-1)$ if $g\leq 0$. However, for positive $g$ it
is preferable have some fermions with $m_j>1$: this increases the
kinetic energy but allows to decrease the correlation energy. In
particular, for very large positive values of $g$ the correlation
energy will dominate. The minimum value $\cE_{\rm corr} = - gN(N-1)$
of the correlation energy is for $[\lambda] = [1^N]$, but we need all
$m_j$'s to be different to get a non-zero such eigenstate. The minimum
kinetic energy possible then is for $m_j = j$. This gives the total
energy $\cE_{[1^N]}=(2B-g)N(N-1)$, which is lower than $E_{[N]}$ for
$g>B$.  One can show that the latter states are groundstate for
$g>2B$.  The groundstates in the immediate regime $0 <g<2B$ can be
found by solving a variation problem. One finds that they are given
by partitions
\eq \lambda_i = \mbox{ non-negative integer close to } \; D - K
i ,\quad 0\leq i \leq D/K \eqend
with $K=(2B-g)/g>0$ and $D=\sqrt{2N K}$ (the corresponding Young
tableaux approximates a right triangle with area $N$ and slope
$K$). One thus finds that the QH model has a nontrivial zero
temperature phase diagram with zero temperature phase transitions at
$g=0$ and $g=2B$. The groundstates are Slater states for $g<0$ and
$g>2B$ but are correlated states for $0<g<2B$. All groundstates are
highly degenerated for $1\ll N\ll \Lambda$, and the degeneracy
(entropy) varies significantly in the region $0<g<2B$.

\Sct{Interacting fermion systems}  In this Section I will discuss
several well-known exactly solvable models of interacting fermions
which play an important role in condensed matter physics.  
This discussion suggests an interesting physical interpretation
of the QH model discussed in the previous section. This interpretation
will lead us to propose a general strategy to obtain `interesting'
simplified models for 2D correlated fermions. 

The general class of interacting fermion models which we consider
describe fermions with 1-particle states labeled by a large but finite
number of different quasi-momenta $\vk$ and a spin index $\al$
assuming at most two different values (`at most' since we also discuss
models for spinless fermions). The models are defined by an
Hamiltonian $H=H_0+V$ with the free part
\eq
\label{H0}
H_0= \sum_{\vk,\al} (E^\nd_{\vk} -\mu) c^\dag _{\vk,\al} c^\nd_{\vk,\al} 
\eqend
where $E^\nd_{\vk}$ is the dispersion relation and $\mu$ the chemical
potential. We assume that the interaction is of the following form,  
\eq 
\label{V}
V = 
\sumdash \sum_{\al,\be} 
v^\nd_{\vk_1 \ldots \vk_4} :
c^\dag_{\vk_1,\al} c^\nd_{\vk_2,\al} c^\dag_{\vk_3,\be} c^\nd_{\vk_4,\be} : 
\eqend 
where $v^\nd_{\vk_1 \ldots \vk_4}$ is the interaction potential in Fourier space. 
The $c^{(\dag)}_{\vk,\al}$ are fermion operators obeying the usual
anticommutator relations
\eq 
\label{car} 
\bigl\{ c^\dag _{\vk,\al} , c^\nd_{\vk',\be} \bigr\} =
\delta_{\vk,\vk'}\delta_{\al,\be} \eqend
etc. The model is defined on the Fermion Fock space with a vacuum
$\Omega$ such that $c_{\vk,\al}\Omega=0$ for all $\al$ and $\vk$.  The
primed sum is over all $\vk_1,\ldots,\vk_4$ restricted by momentum
conservation. In simple cases the latter amounts to
$\vk_1-\vk_2+\vk_3-\vk_4=\vnull$, but if the system is defined on a
spatial lattice with finite lattice constant also umklapp processes
are allowed and `$\vnull$' above can be replaced by any reciprocal
lattice vector.

In the following discussion we will symbolically represent the
different interaction terms in \Ref{V} as follows, 
\eq
\label{FigV} 
\figureV  \qquad \quad \eqend
where spin conservation is understood horizontally. This term
represents the process where two fermions with momenta $\vk_2$ and
$\vk_4$ are scattered and thereby obtain new momenta $\vk_1$ and
$\vk_3$, respectively. The Hubbard model on a finite 2D $L\times L$
lattice corresponds to the special case where\footnote{To see that
this is identical with the standard definition of the 2D Hubbard model
\cite{review} use $n_\uparrow n_\downarrow = \; :(n_\uparrow +
n_\downarrow)^2:/2$ and perform a lattice Fourier transform.}
\eqa
\vk=(k_x,k_y),\quad k_{x,y}=(2\pi/L)\times\mbox{integer},\quad   -\pi\leq
k_{x,y}<\pi, 
\eqaend
$\al=\uparrow,\downarrow$, and
\eq 
\label{Hub} 
E_\vk = -2t(\cos k_x + \cos k_y),\quad v^\nd_{\vk_1 \ldots
\vk_4} = \frac{1}{2 L^2} U \sum_{\vK}\delta_{\vk_1-\vk_2+\vk_3-\vk_4,\vK} , 
\eqend
with $t$ and $U>0$ the usual Hubbard constants; the sum is over all
reciprocal lattice vectors $\vK = (K_x,K_y)$ with $K_{x,y} =
2\pi\times$integer.  Thus in the Hubbard interaction all scattering
processes occur and have equal weight.  The number of distinct
1-particle states in this case is $2L^2$, and the Fermion Fock space
is $2^{2L^2}$-dimensional.

We now discuss various simplified models which can be obtained from a
models of the type given above. Our method is by truncation, leaving
only particular interaction terms. This truncation method might seem
ad-hoc, and I therefore stress that it only should be regarded as a
simple way to find the {\em structure} of `interesting'
Hamiltonians. Systematic derivations of these Hamiltonians should be
based on renormalization group methods; see e.g.\ \cite{Salmhofer} and
references therein.

\sct{Hartree-Fock models} As is well-known, not all interaction terms
in an interacting fermion model are equally important. For example,
the terms where the particle momenta are conserved, $\vk_1=\vk_2$ and
$\vk_3=\vk_4$, or exchanged, $\vk_1=\vk_4$ and $\vk_3=\vk_2$, are
known to be of particular importance. We will refer to them as as
Hartree- and Fock terms, respectively. They can be represented
symbolically as follows,
\eq
\label{FigHF} 
\figureHF \qquad \qquad \quad 
\eqend
\bigskip

\noindent The simplest example of an exactly solvable model is
obtained by truncating the interaction in \Ref{V} and keeping only the
Hartree-terms,
\eq
\label{Hartreemodel} 
\VH = \sum_{\vk,\vq,\al,\be} W_{\vk,\vq} : \hat n^\nd_{\vk,\al} \hat
n^\nd_{\vq,\be} : , \eqend
where I introduced the fermion number operators
\eq
\label{n}
\hat n^\nd_{\vk,\al}\equiv c^\dag _{\vk,\al} c^\nd_{\vk,\al} . 
\eqend
These operators all commute, and thus the Hartree Hamiltonians
$H=H_0+\VH$ are sums of commuting terms and can be easily
diagonalized.  Note that one obtains such a model also if one also
keeps parts of the Fock-terms, namely those with $\al=\be$: this adds
to the interaction the terms
\eq
\label{Fockmodel} 
\VF = -\sum_{\vk,\vq,\al} W_{\vk,\vq} : \hat n^\nd_{\vk,\al} \hat
n^\nd_{\vq,\al} : , 
\eqend
and all we say below can be easily generalized to Hartree-type models
with Hamiltonians $H=H_0 + \VH + \VF$.

We naively described the procedure yielding the simplified Hamiltonian
$H=H_0 + \VH$ from the one in Eq.\ \Ref{H0}--\Ref{V} as truncation
dropping all but certain particular terms in the interaction. This
suggests simple formulas of the parameters $W_{\vk,\vq}$ in terms of
the $v_{\vk_1,\ldots,\vk_4}$. However, a systematic derivation should
be based on some renormalization procedure in which the model
parameters are renormalized and all but certain interaction terms
become irrelevant. Thus this truncation should also be accompanied by
some change of model parameters. One thus should interpret the
Hamiltonian $H=H_0+\VH$ as a phenomenological model with model
parameters to be fixed by experimental data. Similar remarks apply to
all other truncated models discussed below.

The Hartree Hamiltonian $H=H_0+\VH$ is exactly solvable:
as can be checked easily, its exact energy eigenstates are given by
\eq
\label{Psin} 
\Psi_{\bf \vn} = \prod_{\vk,\al} \left( c^\dag_{\vk,\al}
\right)^{n_{\vk,\al}}\Omega , \eqend
and are labeled by the occupation numbers $\vn = \{ n_{\vk,\al}\}$,
$n_{\vk,\al}=0,1$. The corresponding energy eigenvalues are
\eq \label{En} 
\cE_\vn = \sum_{\vk,\al} (E^\nd_{\vk}-\mu ) n^\nd_{\vk,\al} +
\sum_{\vk\neq \vq,\al,\be} W_{\vk,\vq} n^\nd_{\vk,\al}
n^\nd_{\vq,\be} .  \eqend
This exact solution makes the analysis of this model much simpler as
compared to a general interacting fermion model in Eq.\
\Ref{H0}-\Ref{car}, but it important to note that this solution still
is quite far from understanding the physics of this model: to find the
ground state of the Hamiltonian in \Ref{Hartreemodel} at fixed
particle number $N$ one needs to minimize the functional in Eq.\
\Ref{En} over all configurations $\vn$ such that $\sum_{\vk,\al}
n_{\vk,\al}=N$, and even though the exact solutions yields the
following formula for the partition function, 
\eq
\cZ = \sum_{\vn} \ee^{-\beta \cE_\vn} , 
\eqend
its computation still is a non-trivial task (it is equivalent to
solving and Ising-type model). It is interesting to note that such an
analysis of a Hartree-type model appears to be very similar to
Landau's Fermi liquid theory \cite{Landau}.

A important feature of the Hartree-type models is that their
eigenstates are identical with the ones of the corresponding
non-interacting model: the interaction only affects the energy
eigenvalues, but the eigenstates remain Slater states without any
fermion correlations. A more complicated model taking into account
some correlations is obtained if one keeps, in addition to Hartree
terms, all Fock terms (see \Ref{FigHF} above). This yields an
interaction of the following type,
\eq \VHF = \sum_{\vk,\vq} : W_{\vk,\vq} \hat n_\vk \hat
n_\vq + J_{\vk,\vq}\hat \vS_\vk\cdot \hat \vS_\vq : \eqend
with the usual spin- and charge density operators $\hat \vS = (\hat
S^1,\hat S^2,\hat S^3)$ and $\hat n$,
\eq
\hat \vS_\vk = \sum_{\al,\beta} c^\dag_{\vk,\al}
(\vsig)_{\al\be} c_{\vk,\be} \quad \mbox{ and } \quad \hat n_\vk
= \sum_\al c^\dag_{\vk,\al} c^\nd_{\vk,\al} ,
\eqend
and $\vsig=(\sig^1,\sig^2,\sig^3)$ the Pauli spin matrices as
usual.

The Hamiltonian $H=H_0 + \VHF$ is no longer the sum of mutually
commuting terms. Its eigenfunctions thus are more complicated. To see
that it is convenient to write the Slater states in Eq.\ \Ref{Psin} as
follows,
\eq
|\al_1,\ldots, \al_N \rangle = 
c^\dag_{\vk_1,\al_1}\cdots c^\dag_{\vk_N,\al_N}\Omega . 
\eqend
The action of $H=H_0 + \VHF$ on these states is easily computed,
\eq H|\al_1,\ldots, \al_N \rangle = \left( \cE_0 + \sum_{j\neq k}
J(\vk_j,\vk_k) \hat \vsig_j\cdot \hat \vsig_k \right) |\al_1,\ldots,
\al_N \rangle \eqend
with $\cE_0 = \sum_j (E^\nd_{\vk_j}-\mu ) + \sum_{j\neq k}
W_{\vk_j,\vk_j}$ the Hartree-model eigenvalues and the $\hat \vsig_j$
Heisenberg spin operators acting on the spin $\al_j$-index of the
fermions as usual. To find the eigenvalues of the Hartree-Fock model
one thus needs to diagonalize the spin Hamiltonians $\sum_{j\neq k}
J(\vk_j,\vk_k) \hat \vsig_j\cdot \hat \vsig_k$.  The latter can be
interpreted as 1D Heisenberg spin systems with long-range
interactions.  In the case where $J(\vk,\vq)=J$ is constant, the spin
Hamiltonian reduces to $J \left(\sum_{j=1}^N\hat \vsig_k\right)^2 +
const.$, and its eigenvalues and eigenstates can be determined using
group theory (i.e.\ decomposing the direct sum of $N$ spin-$1/2$
representations of the Lie algebra of $\SU(2)$ in irreps).

\sct{BCS-type models} These models underlie the theory of of
superconductivity. They can formally be obtained from a general
interacting fermion model in Eq.\ \Ref{H0}-\Ref{car} by dropping all
but the following terms in the interaction,
\eq
\label{FigBCS}
\figureBCS\qquad \qquad \qquad 
\eqend

\bigskip\noindent The resulting Hamiltonian is $H=H_0 + \VB$ with 
\eq
\VB =  \sum_{\vk,\vq} g_{\vk,\vq} \Delta^+_\vk \Delta^-_\vq 
\eqend
where 
\eq
\Delta^+_\vk = c^\dag_{\vk,\uparrow} c^\dag_{-\vk,\downarrow},\quad 
\Delta^-_\vq = c^\nd_{-\vq,\downarrow} c^\nd_{\vq,\uparrow} . 
\eqend
This Hamiltonian is exactly solvable for parity invariant systems
where $E_\vk=E_{-\vk}$. One way to see this is to write $H_0 =
\sum_{\vk} (E_{\vk}-\mu) \tilde S_\vk$ (up to an irrelevant additive constraint) where
\eq \tilde S_\vk = c^\dag_{\vk,\uparrow} c^\nd_{\vk,\uparrow} -
 c^\nd_{-\vk,\downarrow} c^\dag_{-\vk,\downarrow} .
\eqend
One can check that the operators $\Delta^\pm$ and $\tilde S$ represent
the Lie algebra of $\SU(2)$,
\eq [\Delta^+_\vk,\Delta^-_\vq] = \delta_{\vk,\vq} \tilde S_\vk,\quad
[\tilde S_\vk,\Delta^\pm_q] = \pm \delta_{\vk,\vq} \Delta^\pm_\vk . 
\eqend
Thus the BCS Hamiltonian is a quantum $\SU(2)$ spin-type model of
similar to the Hartree-Fock model described above. One can make this
relation more specific by introducing the following Bogoliubov
transformation,
\eq
\tilde c^\nd_{\vk,\uparrow} \equiv c^\nd_{\vk,\uparrow},\quad 
\tilde c^\nd_{\vk,\downarrow} \equiv c^\dag_{-\vk,\downarrow}
\eqend
leaving the canonical anticommutator relations invariant. Then
$\Delta^\pm_\vk = (\tilde S^1_\vk \pm \ii \tilde S^2_\vk)/2$ and $\tilde
S_\vk=\tilde S_\vk^3$ with $\tilde\vS_\vk = \tilde c_\vk^\dag \vsig
\tilde c^\nd_\vk$ the spin operators associated with these new fermion
operators. The BCS Hamiltonian can then be written as 
\eq H = \sum_{\vk} (E_\vk -\mu)\tilde S^3_\vk + \sum_{\vk,\vq}
g_{\vk,\vq} \left( \tilde S^1_\vk \tilde S^1_\vq +
\tilde S^2_\vk \tilde S^2_\vq \right) \eqend
which formally is identical with a $xy$ Heisenberg spin model in a
transverse magnetic field.  This model can therefore be treated in a
similar manner as the Hartree-Fock model discussed above.  The exactly
solvable case of constant coupling, $g_{\vk,\vq}=g$, underlies the BCS
theory of superconductivity \cite{BCSorig} and was solved exactly in
\cite{BCS}.

\sct{A different interpretation of the QH model} \label{s22} Our
discussion above suggests the following interpretation the interaction
of the QH model: we can regard the quantum numbers $\ell$ and $m$ as
the $x$- and $y$-components of a 2D pseudo momentum, $\ell=k_x$ and
$m=k_y$. We then can interpret $\VQH$ in Eq.\ \Ref{VQH1} as mixed
scattering terms which are Hartree-like in the $x$-component and
Fock-like in the $y$-component of the momenta or {\em vice versa},
i.e.\
\eq
\label{Figmx} 
\figuremx \qquad \qquad \quad 
\eqend

\bigskip\noindent As discussed, one can exactly solve the generalized
QH model $H=H_0 + \VQH+\VH$ defined in Eqs.\ \Ref{H011}, \Ref{VQH1}
and \Ref{VQH2}. Interpreting $E_m +\tilde E_\ell\equiv E_\vk$ as
dispersion relation we can regard this as simplified model obtained by
truncating a spin-less variant of a model defined in Eqs.\
\Ref{H0}--\Ref{V}, keeping not only Hartree- and Fock terms but also
the mixed terms just described.  One main message of this paper is
that {\em it is possible to keep these mixed terms and still have an
exactly solvable model.}

As is well-known, spatial correlations are important in 2D, and the
simplified models models including only the Hartree- and Fock terms
\Ref{FigHF} are not adequate. The main suggestion is {\em to use the
generalized models in 2D where, in addition to the Hartree-Fock terms
\Ref{FigHF}, also the mixed terms \Ref{Figmx} are included}.

It is important to note that, in our present interpretation of the QH
model, interchanging $k_x\leftrightarrow k_y$ amounts to changing
$E_\ell \leftrightarrow \tilde E_\ell$ and $g\leftrightarrow -g$; see
Eqs.\ \Ref{rho}--\Ref{123}. Thus, for spin-less fermions, one can have
a non-trivial correlation interaction $\VQH$ only if parity
invariance is broken. As discussed below, this no longer is the case
for fermions with spin.

\sct{Generalized Hartree model} \label{93} I now define a parity
invariant variant of the QH model including spin. It corresponds to a
generalization of a Hartree model $H_0 + \VH$ defined in Eqs.\
\Ref{H0} and \Ref{Hartreemodel}. We restrict ourselves to 2D
dispersion relation of the following form, $E_\vk = e_{k_x} +
e_{k_y}$, and assume that the Hartree coupling can be written as
$W_{\vk,\vq}= v_{k_x q_x} + v_{k_y q_y} + w_{k_x q_y} + w_{q_x k_y}$.
Introducing the operators
\eq
\label{rr} 
\rho^\nd_{k_y q_y } = \sum_{k_x,\al} c^\dag_{k_x k_y \al}c^\nd_{k_x q_y \al} , \quad 
\tilde \rho^\nd_{k_y q_y} = \sum_{k_y,\al} c^\dag_{k_x k_y \al}c^\nd_{q_x k_y \al} .
\eqend
we can write 
\eq
H_0 = \sum_k e_{k}\bigl( \rho_{k k } + \tilde
\rho_{k k} -\mu\bigr)
\eqend
and 
\eq \VH =  \sum_{k,q} v_{k,q} \bigl(
\rho_{k k} \rho_{q q}+  \tilde \rho_{k k}\tilde
\rho_{q q } ) + w_{k,q} \bigl( \rho_{k k} \tilde \rho_{q
q}+ \tilde \rho_{k k} \rho_{q q } ) .  \eqend
I now propose the model $H=H_0+\VH+\Vmx$ with the following mixed interaction, 
\eq
\Vmx = \sum_{k,q} g :\bigl( \rho_{k q } \rho_{q k} + 
\tilde \rho_{k q} \tilde \rho_{q k}:
\bigr).  
\eqend
This model is obviously parity invariant, and due to the presence of
spin the interaction $\Vmx$ is non-trivial.  It is straightforward to
generalize the discussion in \ref{91} and \ref{92}. One finds that
this model also has a dynamical symmetry $\gl_\infty\oplus\gl_\oplus$,
and explicit formulas for its eigenstates and eigenvalues can be
obtained as well. It would be interesting to explore the physics of
this model in more details.

\sct{On the 2D Hubbard model} To illustrate the flexibility of the
truncation procedure advocated above, I now discuss various simplified
models for the 2D Hubbard model. I also show that the exact solution
of the Hartree-like models for the 2D Hubbard model reproduce mean
field theory, as claimed in the introduction.

\ssct{Simplified models} For the 2D Hubbard model the Hartree-like
model $H_1=H_0 + \VH + \VF$ defined in Eqs.\ \Ref{H0},
\Ref{Hartreemodel} and \Ref{Fockmodel} is given by
\eq
H_0 =  \sum_{\vk} E_\vk\hat n_{\vk}, \qquad 
\VH + \VF= \frac{U}{L^2} \hat n_\uparrow \hat n_\downarrow 
\eqend
with $E_\vk$ in \Ref{Hub} and $\hat n_{\al} = \sum_{\vk} \hat
n_{\vk,\al}$. As I show below, this model is equivalent to mean field
theory (= Hartree-Fock theory) restricted to {\em ferromagnetic} (F)
states. It misses an important physical property of the 2D Hubbard
model which, at half filling, has anti-ferromagnetic (AF) rather than
F order; see e.g.\ \cite{review}. It is interesting to note that there
is also an exactly solvable model accounting for AF: It it obtained by
including Hartree-like terms where the momenta are not fully conserved
but changed by the AF vector $\vQ=(\pi,\pi)$, $\vk_1=\vk_2+\vQ$ and
$\vk_3=\vk_4+\vQ$, and similar Fock-like terms.\footnote{These are
umklapp processes included in the Hubbard interaction in Eq.\
\Ref{Hub}.}  These yields the following interaction term,
\eq \VH^{\vQ} + \VF^{\vQ} 
= \frac{U}{L^2} \hat n^{\vQ}_\uparrow \hat n^{\vQ}_\downarrow , \quad 
n^{\vQ}_\al = \sum_{\vk} c^\dag_{\vk+\vQ,\al}c^\nd_{\vk,\al} . 
\eqend
The model $H_2=H_1+\VH^{\vQ} + \VF^{\vQ}$ can be solved exactly, and its solution
yields Hartree-Fock theory for the 2D Hubbard model allowing for F,
AF, and charge-density waves. 

The corresponding mixed terms are
\eq \Vmx = \frac{U}{L^2}  \sum_{k,q} : \rho_{k q \uparrow} \rho_{q k \uparrow} +
\tilde \rho_{k q \uparrow} \tilde \rho_{q k \uparrow} :  \eqend
with $\rho_{k q}$ and $\tilde \rho_{k q}$ in \Ref{rr}, and similarly
$\Vmx^{\vQ}$. It would be interesting to determine the phase diagram
of the model $H_3 = H_2 + \Vmx + \Vmx^{\vQ}$. I hope to come back to
that in future work.

\ssct{Relation to mean field theory} \label{mft} I now show how to
compute the partition functions of the Hartree-like models above, and
that this reproduces standard mean field theory. For simplicity I
restrict myself to the Hartree-like model allowing for AF only,
\eq
H=H_0 + \frac U{2L^2} (\hat n_\uparrow + \hat n_\downarrow)^2 + \frac U{2L^2}  (\hat n_\uparrow^{\vQ} - 
\hat n_\downarrow^{\vQ} )^2  
\eqend
(we can ignore normal ordering here since $\hat n_\sig^2 = \hat
n^\nd_\sig$, and thus normal ordering only amounts to a shift in the
chemical potential).  It is easy to extend my argument to the other
models discussed above.  We use the following identity
(Hubbard-Stratonovich transformation),
$$ 
\ee^{-\beta H } = const.\
\int_{\R^2} \dd r \dd s\, \ee^{ - \beta \bigl(
L^2(r^2 + s^2)/U + H_0 + s (\hat n^{\vQ}_\uparrow -\hat n^{\vQ}_\downarrow) + 
\ii r(\hat n_\uparrow + \hat n_\downarrow) \bigr)} , $$
which linearizes the interaction at the cost of introducing two
integrations. With that we can compute the partition function (we
drop the irrelevant constant),
\eq 
\label{ccZ}
\cZ = \int_{\R^2} \dd r\dd s\, \ee^{-\beta L^2
[(r^2+s^2)/U - F(r,s)] } \eqend
with $e^{-\beta L^2 F}$ 
the partition function of non-interacting fermions coupled to constant
external fields $r$ and $s$. The latter can be easily computed, and one obtains 
\eq F = \sum_{\al=\pm } \frac1{L^2} \sum_\vk \frac{\dd^2\vk}{(2\pi)^2}
\log(1+\ee^{-\beta \eps_{\vk,\al}})
\eqend
with 
\eq
\eps_{\vk,\pm} = 
(E_\vk + E_{\vk+\vQ})/2 -\mu \pm \sqrt{s^2  + (E_\vk - E_{\vk+\vQ})^2/4}
\eqend
the fermion bands in an AF background. Note that $L^{-2}\sum_\vk$ is a
Riemann sum converging to an integral in the thermodynamic limit
$L\to\infty$. In this limit a saddle point evaluation becomes
exact. The saddle point equations one thus obtains are identical with
Hartree-Fock theory for the Hubbard model restricted to AF states; see
e.g.\ \cite{review}. 

It is interesting to note that Eq.\ \Ref{ccZ} was obtained by a
different method in Ref.\ \cite{LW}. In this work we proposed it as a
useful alternative formulation of mean field theory for the 2D Hubbard
model. Superficially this seems equivalent to the `standard'
formulation which only uses the saddle point equations resulting from
that integral, but it is in fact more general since it allows for the
possibilities of having degenerate saddle points. We found that
degenerate saddle points occur in a large part of the parameter
regime: if one fixes the fermion density close but away from half
filling one typically needs to adjust the chemical potential $\mu$
such that the integral is dominated by {\em two} distinct saddle
points at the same time, one describing AF order, and another
describing no order at all \cite{LW}. These mixed phases are missed if
one only looks at the saddle point equations: the mean field phase
diagram of the 2D Hubbard model thus obtained is much richer than
generally believed (see e.g.\ \cite{review} and references
therein). Mixed phases are typical in Hubbard-like models. I feel this
is a basic, important property of these models which deserves to be
wider known. Anyway, the present paper gives models for which the
mixed phases are exact.

\bigskip

\Sct{Conclusions} In this paper I advocated the following strategy for
treating difficult many-body Hamiltonians for interacting fermions:
rather than solving such model in some approximations (like mean field
theory), truncate it so as to obtain a simplified model which can be
solved exactly. I recalled examples from condensed matter physics a
long time ago where this strategy was used successfully to solve
important problems like superconductivity or magnetism.  I also
discussed a many-body model for 2D quantum Hall systems proposed and
solved in \cite{EL}.  I then argued that this model is the simplest
example in a large class of similar models which should be useful for
shedding light on an interesting problem in modern condensed matter
physics: to understand 2D correlated fermion systems. As indicated,
the partition function of these generalized models can be represented
as matrix models. The computations of the latter are interesting
problems for the future.

Many of the results reported here can adapted straightforwardly to
boson systems. 

\bigskip

\noindent {\bf Acknowledgements:} I thank H.\ Grosse, M.\ Gulacsi, J.\
Mickelsson, M.\ Salmhofer. M.\ Wallin, and R.\ Wyss for useful
discussions and suggestions. I thank S. Rydh for collaboration on
deriving the results described in Section \ref{LR}. This work was
supported by the Swedish Science Research Council (VR) and the G\"oran
Gustafssons Foundation.


\bigskip

\begin{thebibliography}{99}

\bibitem{EL} E.~Langmann, {\em Nucl.\ Phys.\ B} {\bf 654}, 404 (2003)
[arXiv:hep-th/0205287]

\bibitem{review} M.~Imada, A.~Fujimori, and Y.~Tokura, {\em
Rev.\ Mod.\ Phys.} {\bf 70}, 1039 (1998)


\bibitem{Mehta} M.~L.~Mehta, `Random matrices', Academic Press (1991) 

\bibitem{NuclPhys} for review see e.g.\ A.\ Klein and E.\ R.\
Marshalek, {\em Rev.\ Mod.\ Phys.} {\bf 63}, 375 (1991)


\bibitem{BCSorig} J.~Bardeen, L.~N.~Cooper, and J.~R.~Schrieffer, 
{\em Phys. Rev.} {\bf 108}, 1175 (1957) 

\bibitem{BCS} R.\ W.\ Richardson, {\it Phys.\ Lett.} {\bf 3}, 277
(1963); {\it J.\ Math.\ Phys.} {\bf 6}, 1034 (1965)

\bibitem{BCSnew} G.\ Sierra, J.\ Dukelsky, G.\ G.\ Dussel, J.\ von
Delft, and F.\ Braun, {\em Phys.\ Rev. B} {\bf 61}, R11890 (2000); see
also A.\ Mastellone, G.\ Falci, and R.\ Fazio, {\em Phys.\ Rev.\
Lett.} {\bf 80}, 4542 (1998); J.~Links, H.-Q.\ Zhou, R.\ H.\ McKenzie,
M.\ D.\ Gould [cond-mat/0110105]; H.-Q.\ Zhou, J.\ Links, R.\ H.\
McKenzie, and M.\ D.\ Gould, {\it Phys.\ Rev.\ B} {\bf 65}, 060502(R)
(2002)

\bibitem{BCSnew1} L.\ Amico, A.\ Di Lorenzo, and A.\ Osterloh, {\em
Phys.\ Rev.\ Lett.} {\bf 86}, 5759 (2001); J.\ Dukelsky, C.\ Esebbag,
and P.\ Schuck, {\em Phys.\ Rev.\ Lett.} {\bf 87}, 066403 (2001);
X.-W.\ Guan, A.\ Foerster, J.\ Links, H.\-Q.\ Zhou,
{\em Nucl.\ Phys.\ B}{\bf 642}, 501 (2002)

\bibitem{EL0} E.~Langmann, cond-mat/0206045 v1 

\bibitem{LS} E.~Langmann and R.~J.~Szabo, {\em Phys.\ Lett.\ B} {\bf
533}, 168 (2002) [arXiv:hep-th/0202039]

\bibitem{LSZ} E.~Langmann, R.~J.~Szabo and K.~Zarembo,
arXiv:hep-th/0303082

\bibitem{NO} J.~W.~Negele and H.~Orland, `Quantum Many-Particle
Systems', Advanced book classic, Perseus books (1998)


\bibitem{Salmhofer} M. Salmhofer, `Renormalization: An Introduction',
Springer (1999)

\bibitem{Landau} L.~D.~Landau, Sov.\ Phys.\ JETP {\bf 3}, 920 (1957);
{\em ibid.}  {\bf 5}, 101 (1957); {\em ibid.}  {\bf 8}, 70 (1958)

\bibitem{LW} E.~Langmann and M.~Wallin, Phys. Rev. B {\bf 55}, 9439 (1997) 

\end{thebibliography}
\end{document}